\documentclass[aps,prl,showpacs,floatfix,twocolumn,superscriptaddress,byrevtex]{revtex4-1}

\pdfoutput=1

\usepackage{amsmath}
\usepackage{amssymb}
\usepackage{amstext}
\usepackage{amsopn}
\usepackage{amsfonts}
\usepackage{amsxtra}

\usepackage{dcolumn}
\usepackage{graphicx}
\usepackage{bm}

\usepackage{hyperref}
\usepackage{xcolor}
\definecolor{linkcol}{rgb}{0,0,0.4} 
\definecolor{violet}{rgb}{0.32,0,0.52} 
\definecolor{citecol}{rgb}{0.5,0,0} 

\hypersetup{
bookmarksopen=true,
pdftitle="Infrared and Raman spectroscopy measurements of a transition in the crystal structure and a closing of the energy gap of BiTeI under pressure",
pdfauthor="M. K. Tran et al.", 
pdfsubject="Infrared and Raman spectroscopy measurements of a transition in the crystal structure and a closing of the energy gap of BiTeI under pressure", %subject of the document
pdfstartview={FitH},    % fits the width of the page to the window
pdfmenubar=true, %menubar shown
pdfhighlight=/O, %effect of clicking on a link
colorlinks=true, %couleurs sur les liens hypertextes
pdfpagemode=UseNone, %aucun mode de page
pdfpagelayout=SinglePage, %ouverture en simple page
pdffitwindow=true, %pages ouvertes entierement dans toute la fenetre
linkcolor=blue, %couleur des liens hypertextes internes
citecolor=violet, %couleur des liens pour les citations
urlcolor=violet
}

\begin{document}

\title{Infrared and Raman spectroscopy measurements of a transition in the crystal structure and a closing of the energy gap of BiTeI under pressure}

\author{M. K. Tran}
\email{michael.tran@unige.ch}
\author{J. Levallois}
\affiliation{D\'epartement de la Mati\`ere Condens\'ee, University of Geneva, CH-1211 Geneva 4, Switzerland}

\author{P. Lerch}
\affiliation{Swiss Light Source, Paul Scherrer Institute, CH-5232 Villigen, Switzerland}

\author{J. Teyssier}
\author{A. B. Kuzmenko}
\affiliation{D\'epartement de la Mati\`ere Condens\'ee, University of Geneva, CH-1211 Geneva 4, Switzerland}

\author{G. Aut\`es}
\author{O. V. Yazyev}
\affiliation{Institute of Theoretical Physics, Ecole Polytechnique F\'ed\'erale de Lausanne (EPFL), CH-1015 Lausanne,
Switzerland}

\author{A. Ubaldini}
\author{E. Giannini}
\author{D. van der Marel}
\author{A. Akrap}
\email{ana.akrap@unige.ch}
\affiliation{D\'epartement de la Mati\`ere Condens\'ee, University of Geneva, CH-1211 Geneva 4, Switzerland}

\date{\today}

\begin{abstract}
BiTeI is a giant Rashba spin splitting system, in which a non-centro symmetric topological phase has recently been suggested to appear under high pressure. We investigated the optical properties of this compound, reflectivity and transmission, under pressures up to $15$~GPa. The gap feature in the optical conductivity vanishes above $p \sim 9$~GPa and does not reappear up to at least $15$~GPa. The plasma edge, associated with intrinsically doped charge carriers, is smeared out through a phase transition at $9$~GPa. Using high pressure Raman spectroscopy, we follow the vibrational modes of BiTeI, providing additional clear evidence that the transition at 9 GPa involves a change of crystal structure. This change of crystal structure possibly inhibits the high-pressure topological phase from occurring.
\end{abstract}


\maketitle

Interest in the non-centrosymmetric semiconductor BiTeI surged when it was found that this compound hosts the largest known Rashba spin splitting in bulk form~\cite{ishizaka11, Crepaldi12, Landolt12}. While this material is structurally related to the recently discovered bismuth chalcogenide topological insulators~\cite{Xia09,Zhang09}, it is an insulator of the common variety at ambient pressure. Recent first-principles band structure calculations suggested that BiTeI undergoes a transition to the topological insulating phase under pressure~\cite{bahramy12}, through which BiTeI would become the first example of non-centrosymmetric topological insulator. Moreover, such a band-structure topology change realizes a remarkable example of topological phase transition. While several examples of topological phase transitions occurring upon varying chemical composition have been reported in the literature~\cite{Xu11,Sato11,Brahlek12}, the pressure-induced transition in BiTeI would present the advantage of being controllable and reversible. 

Optical conductivity is well suited to probe the band structure of BiTeI under pressure. In this Letter, we determine the high pressure optical properties by measuring transmission and reflectivity of BiTeI up to $15$~GPa. We follow the optical gap under pressure and find that it decreases monotonically until $9$~GPa. At this pressure the plasma edge associated with the doped carriers is strongly broadened due to a sudden increase of $\sigma_1(\omega)$ at the plasma frequency. Above this pressure the gap feature in the optical conductivity has disappeared, and it does not reappear to the highest pressure reached. The high pressure phase appears to be metallic. Using Raman spectroscopy, we observe a sudden change in the number and frequency of the vibrational modes at $9$~GPa, which shows that a structural transition occurs at this pressure. 

Single crystals of BiTeI were grown by the floating zone method, starting from the stoichiometric ratio of metallic bismuth, tellurium and bismuth iodide. The unit cell of BiTeI is composed of  triple layers, Te--Bi--I, stacked along the polar {\em c}-axis~\cite{ishizaka11}. The triple layers are bound by a weak van der Waals interaction. The structure is described by the non-centrosymmetric space group $P3m1$.

At ambient pressure, the reflectance was measured at a near-normal angle of incidence on a freshly-cleaved $a-b$ surface, from $25$~meV to $0.9$~eV using an {\em in situ} evaporation technique. The complex dielectric function and the optical conductivity were obtained by Kramers-Kronig transformation of the reflectivity, supported by spectroscopic ellipsometry data from  $0.6$~eV to $5.6$~eV. Knowing the optical constants of diamond, we converted the vacuum-sample reflectivity to one at the diamond-sample interface, $R_{sd}$. Transmission was measured through a flake of BiTeI using an infrared microscope. Optical data at high pressure were acquired up to $15$~GPa with a diamond anvil cell (DAC) at the X01DC IR synchrotron beamline of the Swiss Light Source~\cite{Akrap12, lerch12}. Reflectivity was determined in the frequency range from $60$~meV to $1$~eV for a dense set of pressures. The intensity of radiation reflected from the diamond to sample interface was normalized by the intensity from the interface between the diamond and the metallic (Cu-Be alloy) DAC gasket. 

The reflectivity ratios measured in the DAC were calibrated against $R_{sd}$ obtained from the absolute reflectivity measured outside the cell. In order to calculate the complex dielectric function $\hat{\epsilon}(\omega)$  the reflectivity data were fitted for all pressures to a Drude-Lorentz expression. A sufficiently large number of oscillators (variational dielectric function~\cite{kuzmenko}) was used so as to reproduce all fine details of the reflectivity spectra. Transmission was measured from $80$~meV to $1$~eV for a fine mesh of pressure points. Raman spectra at high pressure were recorded using the same DAC up to $15$~GPa, with daphne oil $7373$ as pressure medium and a home-made micro Raman spectrometer.

\begin{figure}[tb]
\centering
\includegraphics[width=\columnwidth]{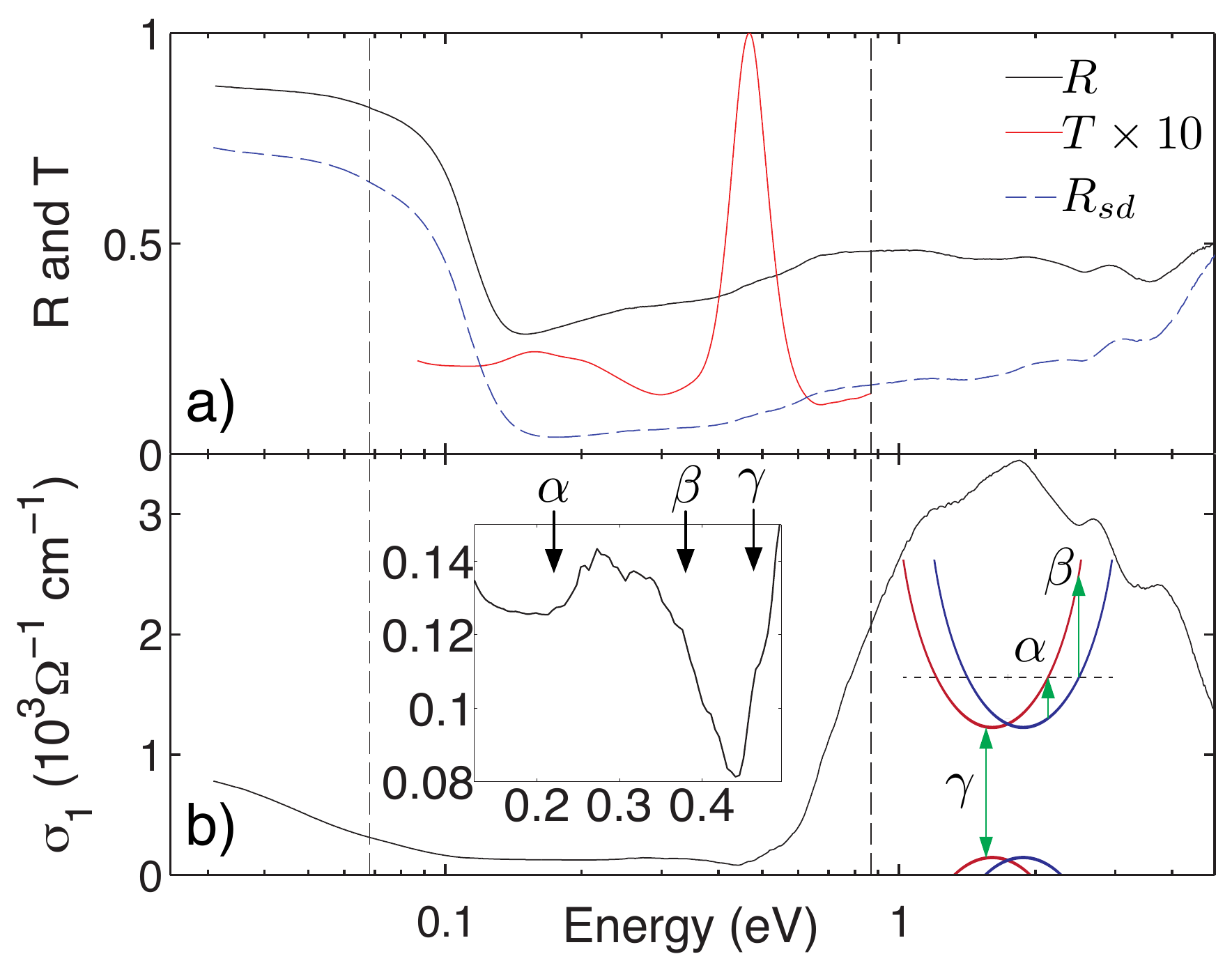}
\caption{(Color online) (a) The reflectance and transmission, and (b) the real part of the complex conductivity $\sigma_1(\omega)$. The left inset: blow-up of $\sigma_1(\omega)$. The right inset: sketch of band structure near the $A$ point with the chemical potential in dashed line. The dashed vertical lines mark the energy range where the high-pressure experiments were perfomed.}
\label{fig1}
\end{figure}

Reflectance and transmission data are shown in Fig.~\ref{fig1}a at ambient pressure and room temperature. A sharp plasma edge appears in the reflectance at $0.15$~eV, and near $0.4$~eV a kink is observed. Transmission through a $6~\mu$m thick flake of BiTeI in the mid infrared range displays a maximum around $0.45$~eV, which matches the frequency of the kink in the reflectivity. Another small peak in the transmission is observed at $0.15$~eV and concurs with the plasma edge in the reflectance. $R_{sd}$ is the reflectivity expected for the diamond-sample interface, showing similar features as the reflectivity with a plasma edge at the same position.

The real part of the optical conductivity $\sigma_1(\omega)$ shown in Fig.~\ref{fig1}b is gapped below $\sim 0.5$~eV, which is just above the energy of the maximum in transmission. Simultaneously, a Drude contribution emerges below 0.1~eV which agrees with the metallic behavior of the resistivity as well as with the previous optical studies~\cite{lee11,martin12,demko12}. This metallic contribution is very likely due to the impurity doping and a slight off-stoichiometry, particularly of iodine because of its high fugacity. We attribute the fine structures in the optical conductivity to the interband transitions $\alpha$ and $\beta$, and the onset of absorption (gap edge) $\gamma$, as sketched in the insets of Fig.~\ref{fig1}b ~\cite{lee11,martin12, demko12}. 

\begin{figure*}[tb]
\centering
\includegraphics[width=1.0\textwidth]{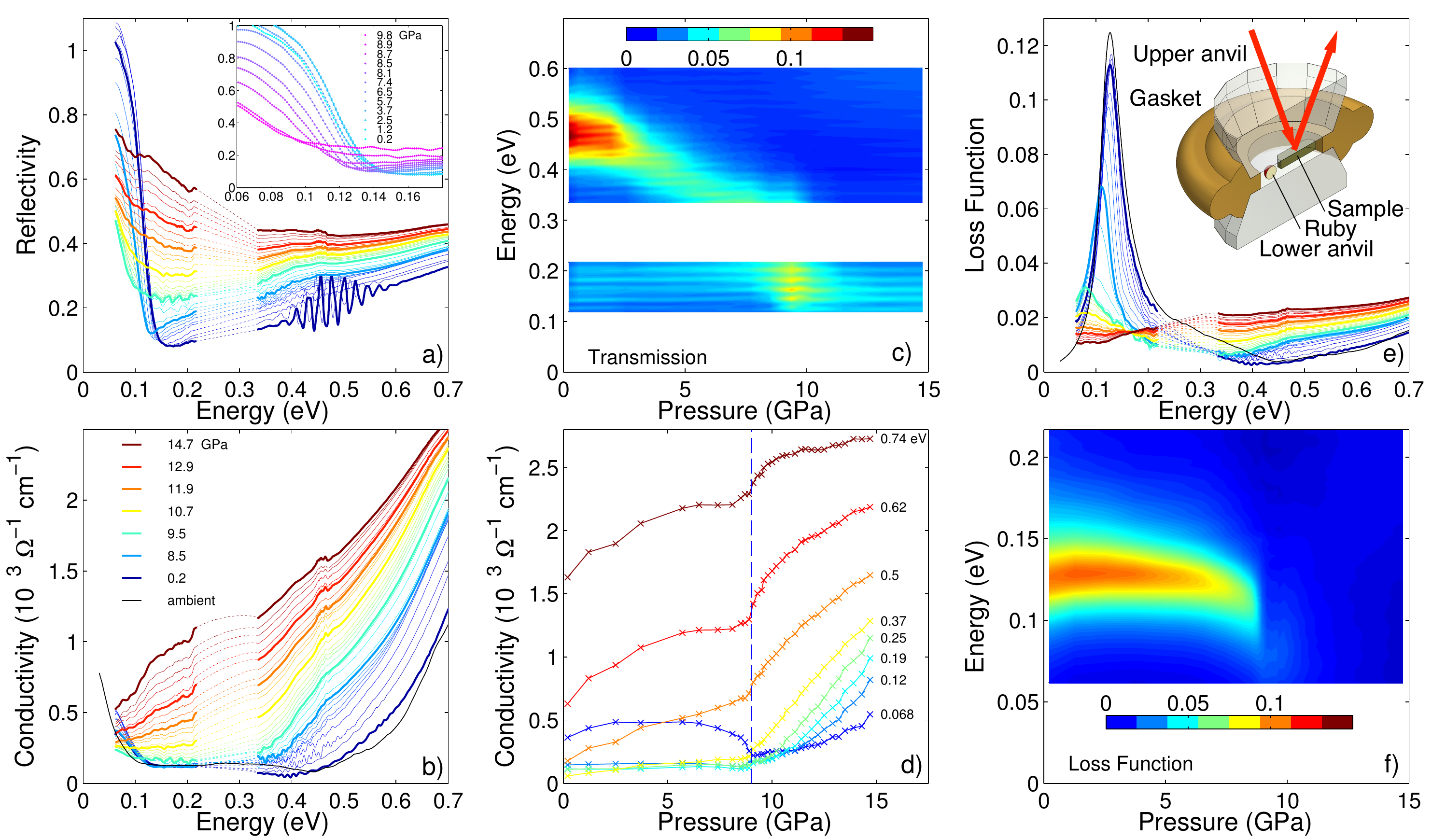}
\caption{(Color online) (a) Reflectivity shown for different pressures (legend in panel~b). Inset: Close up in the low pressure range. (b)~Real part of the optical conductivity, $\sigma_1(\omega)$. (c)~Transmission through a $\sim 5\mu$m thick sample. (d)~The pressure dependence of $\sigma_1(\omega)$ at different energies. (e)~and (f) The loss function, $-\text{Im} (1/\hat{\epsilon}(\omega))$. Inset shows a sketch of the sample environment within the diamond anvil pressure cell. In all panels, the data taken between $0.22$ and $0.33$~eV are not shown due to strong diamond absorption.}
\label{fig2}
\end{figure*}

The band structure and the interband transitions are expected to be strongly influenced by pressure~\cite{bahramy12}. Indeed, we observe significant pressure induced changes in the optical properties. Fig.~\ref{fig2}a shows the reflectivity as a function of photon energy for various pressures. The reflectivity is measured at a sample-diamond interface, and the levels may be compared to the $R_{sd}$ shown in Fig~\ref{fig1}a. The lowest pressure curves contain Fabry-Perot resonances at frequencies just below the gap onset, due to the small sample thickness and the sample transparence. The reflectivity in Fig.~\ref{fig2}a has a sharp plasma edge at $0.15$~eV, consistent with Fig.~\ref{fig1}a. Very little change is observed up to $7$~GPa, and the plasma edge remains at the same energy. There is, however, a gradual increase of the reflectivity level at frequencies above the plasma edge, and a decrease of the level below the plasma edge. Above $\sim 7$~GPa, the plasma edge suddenly begins to shift towards lower energies, and at $9$~GPa it starts to gradually smooth out. Above $9.5$~GPa, the plasma edge disappears, and the reflectivity level now increases across the entire photon energy region. For the highest pressure reached no plasma edge is observed, but the reflectivity has an upturn towards low energies suggesting a metallic state. Overall, the reflectivity experiences a dramatic pressure-induced change in the studied energy range.

The real part of the optical conductivity $\sigma_1(\omega)$ obtained from the complex dielectric function $\hat{\epsilon}(\omega)$ is shown in Fig.~\ref{fig2}b~\footnote{The Fabry-Perot oscillations present in the reflectivity in the lower pressure range give artifact oscillations in the dielectric function and the related quantities; these oscillations are therefore filtered out.}.
Below $9$~GPa, the Drude contribution persists almost unchanged, and the high-frequency $\sigma_1(\omega)$ indicates a gradual shift of the absorption edge towards lower energies. This means that below 0.2~eV there is little change in $\sigma_1(\omega)$, whereas above $0.35$~eV $\sigma_1(\omega)$ increases as the gapped states become filled. As the absorption edge shifts, the dip in $\sigma_1(\omega)$ at $0.45$~eV gradually disappears with increasing pressure. Above $9$~GPa, the Drude contribution disappears from our energy window and $\sigma_1(\omega)$ becomes rather flat below $0.2$~eV. Around this pressure the slope of $\sigma_1(\omega)$ changes sign at low energies, from negative to positive. As the pressure increases further, the gap edge shifts more rapidly to lower energies. By $12$~GPa, the low energy states (below $0.2$~eV) have been filled and the gap tends toward very low energies. At the highest reached pressures, the gap in $\sigma_1(\omega)$ appears to be zero or very small. There are two clear limiting cases for the optical gap. At zero pressure, the gap is finite ($\sim 0.4$~eV), and remains above $\sim 0.3$~eV up to 9~GPa. At $15$~GPa the conductivity is not gapped at room temperature. We cannot attach a precise value to the gap above $9$~GPa because it is not accompanied by a clear spectroscopic feature.

The transmission, shown in Fig.~\ref{fig2}c~\footnote{\label{TSupp}See also Fig.~4 of the supplemental material}, features a sharp peak at $0.46$~eV at low pressure, just below the band gap. It agrees with the ambient pressure transmission shown in Fig.~\ref{fig1}a. The position of the transmission maximum redshifts with pressure and diminishes in intensity. This decrease in the maximum intensity is particularly sharp at $3$~GPa. The monotonic decrease terminates in an abrupt disappearance of the peak at $9$~GPa. 
The collapse of the transmission peak above $9$~GPa suggests that a phase transition is taking place at this pressure. This is in agreement with the reflectivity and the optical conductivity, which indicate that above $9$~GPa the gap feature in $\sigma_1(\omega)$ is gone.

Fig.~\ref{fig2}d shows $\sigma_1(\omega)$  as a function of pressure, for a series of photon energies ranging from $0.068$ eV to $0.74$ eV. Common to all the curves is a sharp kink at $p\sim 9$~GPa. This confirms that at $9$~GPa a phase transition takes place which influences the electronic structure. Above $9$~GPa, $\sigma_1(p)$ steeply increases in the whole energy range. The optical conductivity $\sigma_1$ at $0.068$~eV shown in Fig.~\ref{fig2}d (corresponding to the lower limit of our experimental window of observation), the optical conductivity spectra shown in Fig.~\ref{fig2}b and the transmission spectra in Fig.~\ref{fig2}c~\footnotemark[\value{footnote}], have a gradual variation as a function of pressure between $0$ and $9$~GPa. In order for our data to be consistent with a quantum phase transition in the range $1.7 - 4.1$~GPa as predicted in Ref.~\cite{bahramy12} and the signature of which was reported in Ref.~\cite{xi13}, it would be necessary that the full process of closing and reopening of the gap is completed in between the pressure points of Fig.~\ref{fig2}, \emph{i.e.} within a pressure window of about 0.5 GPa.

\begin{figure*}[tb]
\centering
\includegraphics[width=1.0\textwidth]{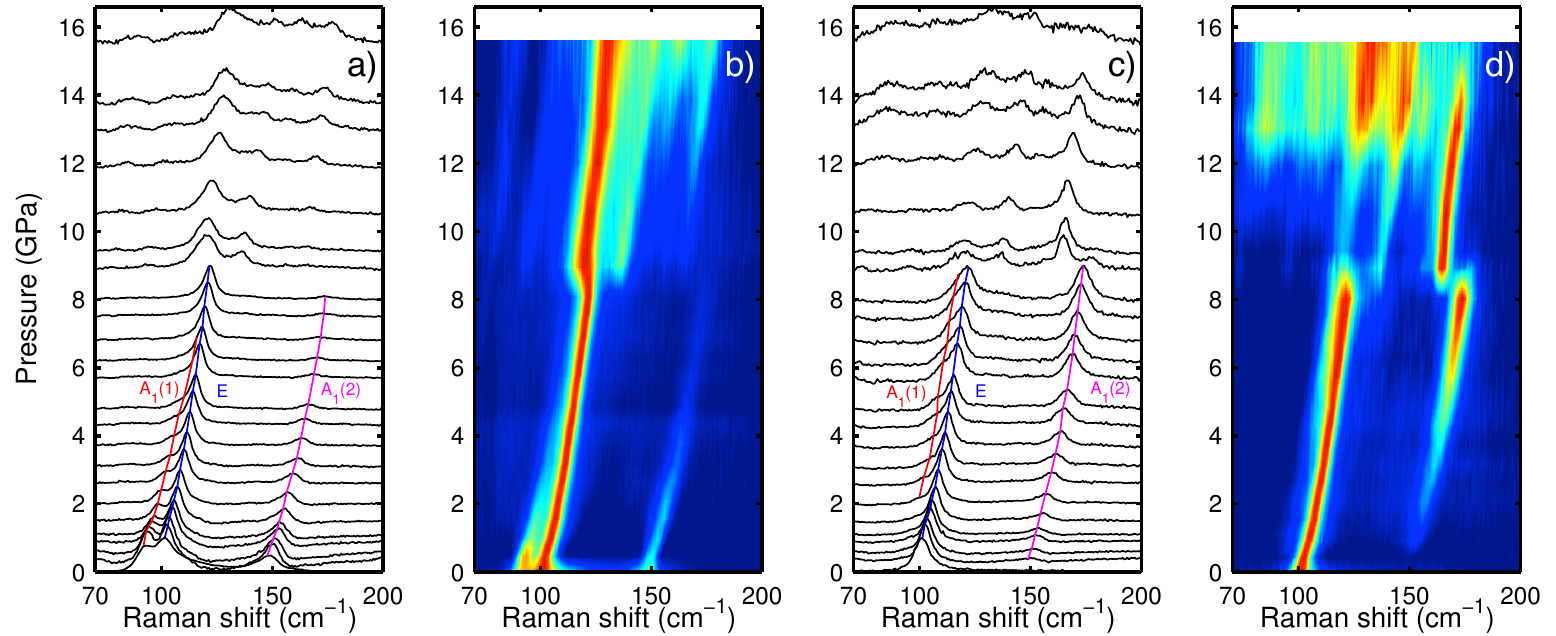}
\caption{(Color online) Raman spectra from 0 to 15~GPa. (a) and (b): parallel polarization. (c) and (d) crossed polarization.} 
\label{fig3}
\end{figure*}

One of the most striking aspects of the presented data is the abrupt disappearance of the plasma edge for $p\simeq 9$~GPa in the reflectivity. The sharp low-pressure plasma edge originates in a well defined Drude peak in $\sigma_1(\omega)$, which suggests a coherent charge transport in the low pressure phase ($p<9$~GPa). To better illustrate the pressure dependence of the plasma frequency, Fig.~\ref{fig2}e and \ref{fig2}f show the loss function, $-\text{Im} (1/\hat{\epsilon})$.  A maximum in the loss function corresponds to the screened plasma frequency. At ambient pressure we observe one such mode at $0.13$~eV and associate it with the intraband plasmon. This plasmon mode widens with pressure and slightly bends towards lower energies above $3$~GPa, without appreciable pressure dependence up to $9$~GPa. The loss function remaining constant means that the spectral weight of the Drude contribution does not change much. However, at $9$~GPa the peak in the loss function suddenly vanishes. The dramatical disappearance of the plasmon peak is not due to the disappearance of the free carriers: there still has to be a Drude contribution at highest pressures, since the reflectivity has clear Hagen-Rubens upturns at low energies. Instead, the conductivity $\sigma_1(\omega)$ increases above $9$~GPa and this sudden increase in the conductivity produces very strong damping. The plasmon thus becomes overdamped by background conductivity and appears washed out. High pressure pushes the interband transitions close to zero energy, and above $9$~GPa they become strongly mixed with the tail of the Drude peak.

To gain better understanding of the transition at $9$~GPa, we measured the pressure effects on the optical phonons using polarized Raman spectroscopy from $0$ to $15$ GPa, shown in Fig.~\ref{fig3}. The symmetry analysis gives four zone-center vibrational modes, with the irreducible vibrational representation $\Gamma = 2A_1+2E$. All of them are both Raman and infrared active because of the lack of inversion symmetry. At ambient pressure, parallel polarization shows $A_1(1)=94$, $A_1(2)=150$ and $E=102$~cm$^{-1}$, while crossed polarization shows $E=102$~cm$^{-1}$, in agreement with the previous studies~\cite{Sklyadneva12,Gnezdilov13}. Around 2~GPa both $A_1$ modes diminish in the parallel polarization and at the same time appear in the crossed polarization. While this may be related to a gradual change in the crystal lattice, there is no theoretical argument why a topological phase transition would have this effect on the Raman spectra. Besides the expected blueshift of the Raman modes under pressure, an important effect is observed above $9$~GPa: three modes are visible up to $9$~GPa, above which the spectrum changes abruptly and several weak peaks appear. The appearance of new vibrational lattice modes points to a change of crystal structure at the pressure where also the change in optical conductivity takes place. 

Bahramy {\it et al.} predicted that pressure causes reduction of the band gap of the topologically trivial BiTeI~\cite{bahramy12}. As pressure reaches its critical value $p_c$ the gap closes developing a Weyl semimetal phase, and its further increase reopens the gap resulting in a topological insulator phase characterized by a band-inversion feature. The first-principles calculations performed at the DFT-GGA level of theory estimate the critical pressure $p_c$ to be in the range $1.7-4.1$~GPa~\cite{bahramy12}. However, it was shown recently that the inclusion of many-body quasiparticle corrections calculated within the {\em GW} approximation results in a significant increase of the band gap of BiTeI bringing it in close agreement with the experimental value $E_g=0.38$~eV~\cite{Rusinov13}. Such an increase of the electronic band gap in the topologically trivial regime would effectively result in a positive shift of $p_c$~\cite{Yazyev12}. Our {\em GW} calculations estimate $p_c\sim10$~GPa~\footnote{See Supplemental Material for the description of DFT and {\em GW} first-principles calculations}, {\em i.e.} the structural phase transition takes place before the predicted topological phase transition and precludes its observation.

The appearance of several additional Raman active modes above the critical pressure is unambiguous evidence for a change of crystal structure, in agreement with the XRD data~\cite{xi13}. Yet the optical conductivity in Fig.~\ref{fig2} progresses gradually through this phase transition as a function of pressure, and has a kink at $9$~GPa for all infrared photon energies. Therefore we postulate that the high pressure phase is structurally close to the one at ambient pressure (space group $P3m1$ with $3$ atoms per unit cell), and differs mainly by the stacking of BiTeI layers. A good candidate is the space group $P6_3mc$ with $6$ atoms per unit cell, which is the structure of the chemically-related compound BiTeCl at ambient pressure~\cite{Shevelkov1995}. Our {\em ab initio} calculations imposing the latter crystal structure for pressurized BiTeI~\footnotemark[\value{footnote}], show that in this phase the gap also decreases gradually as a function of pressure, and evolves toward a topologically trivial semi-metallic state characterized by a negative gap. This hypothesis about the structure is also consistent with the observation that the main Raman lines are only slightly shifted above $9$~GPa and several other weak lines appear. Another candidate which was recently suggested for the higher pressure phase is the $Pnma$ space group structure with 4 formula units per unit cell~\cite{chen2013}.

To summarize, we have determined the pressure dependence of the optical properties and Raman vibrational modes of BiTeI, under pressures up to $15$~GPa. The reflectivity and transmission show dramatic changes under pressure. The band gap monotonically decreases as a function of increasing pressure and the optical conductivity increases. At $9$~GPa this process is accelerated. As a result, the interband plasmon becomes overdamped above $9$~GPa. The transmission data demonstrate clear evidence of a phase transition at $9$~GPa, corroborated by the pressure-dependent Raman spectra indicating a sudden change to a different crystal structure at this pressure. This structural transition possibly prevents the predicted topological high-pressure phase in BiTeI from occurring.

We thank B. Andrei Bernevig for illuminating discussions. We are grateful to Florence L\'evy-Bertrand, Riccardo Tediosi and Mehdi Brandt for technical assistance. Research was supported by the Swiss NSF through Grants Nos. 200020-135085 and NCCR MaNEP. A.A. acknowledges funding from ``Boursi\`eres d'Excellence'' of the University of Geneva. First-principles computations have been performed at the Swiss National Supercomputing Centre (CSCS) under project s443. G.A. and O.V.Y. were supported by the Swiss National Science Foundation (grant No. PP00P2\_133552) and by the ERC Starting Grant ``TopoMat'' (grant No. 306504).

\bibliography{BiTeI}
\end{document}

\preprint{0}
\title{Supplemental Material: Infrared and Raman spectroscopy measurements of a transition in the crystal structure and a closing of the energy gap of BiTeI under pressure}

\author{M. K. Tran}
%\email{michael.tran@unige.ch}
\author{J. Levallois}
\author{J. Teyssier}
\author{A. B. Kuzmenko}
\affiliation{D\'epartement de la Mati\`ere Condens\'ee, University of Geneva, CH-1211 Geneva 4, Switzerland}

\author{G. Aut\`es}
\author{O. V. Yazyev}
\affiliation{Institute of Theoretical Physics, Ecole Polytechnique F\'ed\'erale de Lausanne (EPFL), CH-1015 Lausanne,
Switzerland}

\author{P. Lerch}
\affiliation{Swiss Light Source, Paul Scherrer Institute, CH-5232 Villigen, Switzerland}

\author{A. Ubaldini}
\author{E. Giannini}
\author{D. van der Marel}
\author{A. Akrap}
%\email{ana.akrap@unige.ch}
\affiliation{D\'epartement de la Mati\`ere Condens\'ee, University of Geneva, CH-1211 Geneva 4, Switzerland}

\date{\today}
\maketitle

\section{1. DFT and GW calculations}
\begin{figure}[htb]
\centering
\includegraphics[width=0.8\columnwidth]{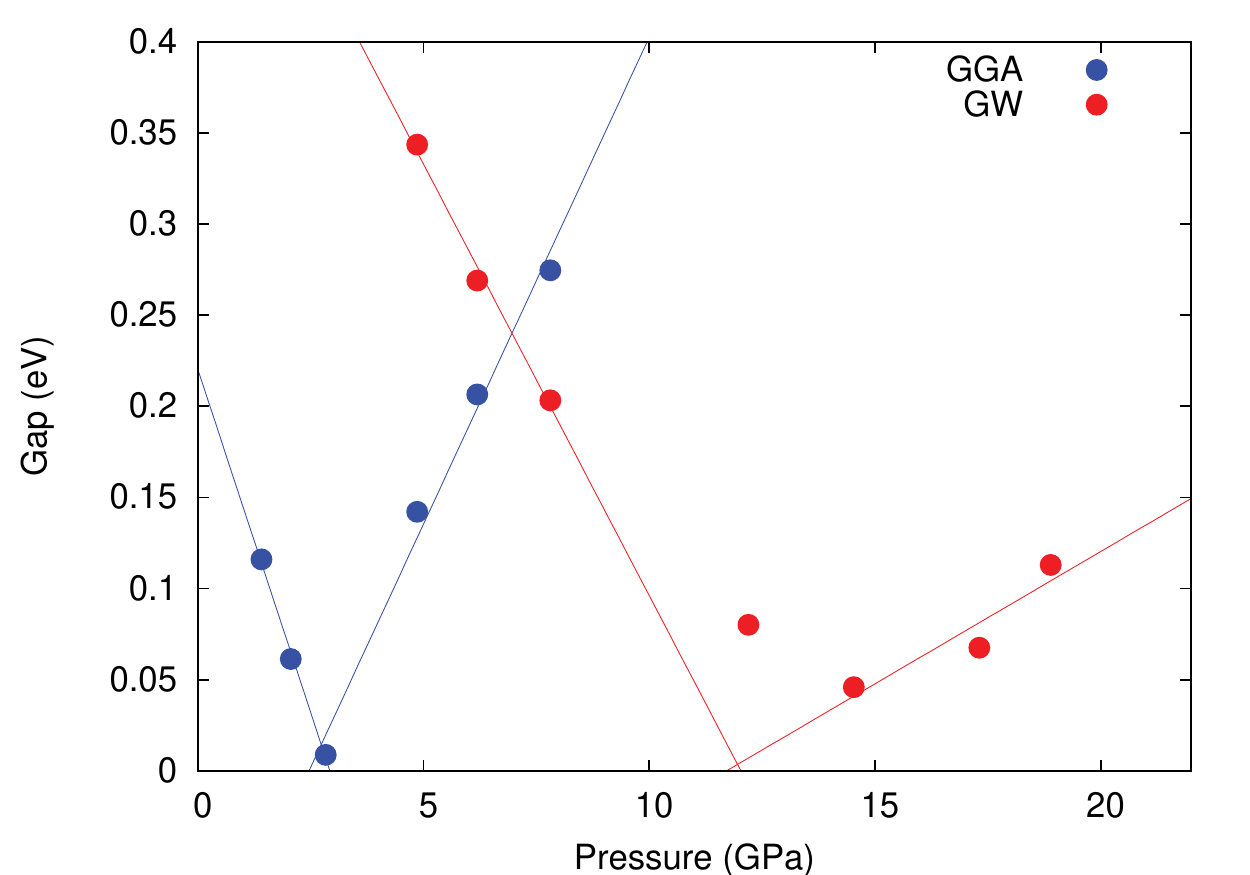}
\caption{Gap as a function of pressure for GGA and GW calculations. }
\label{fig:figs2}
\end{figure}

The band structures of BiTeI under pressure were first calculated from density functional theory (DFT) within the generalized gradient approximation (GGA) as implemented in the QUANTUM-ESPRESSO package~\cite{gia09}. 
Spin-orbit effects were accounted for using the fully relativistic norm-conserving pseudopotentials acting on valence electron wavefunctions represented in the two-component spinor form~\cite{dal05}.

\begin{figure}[tb]
\centering
\includegraphics[width=0.8\columnwidth]{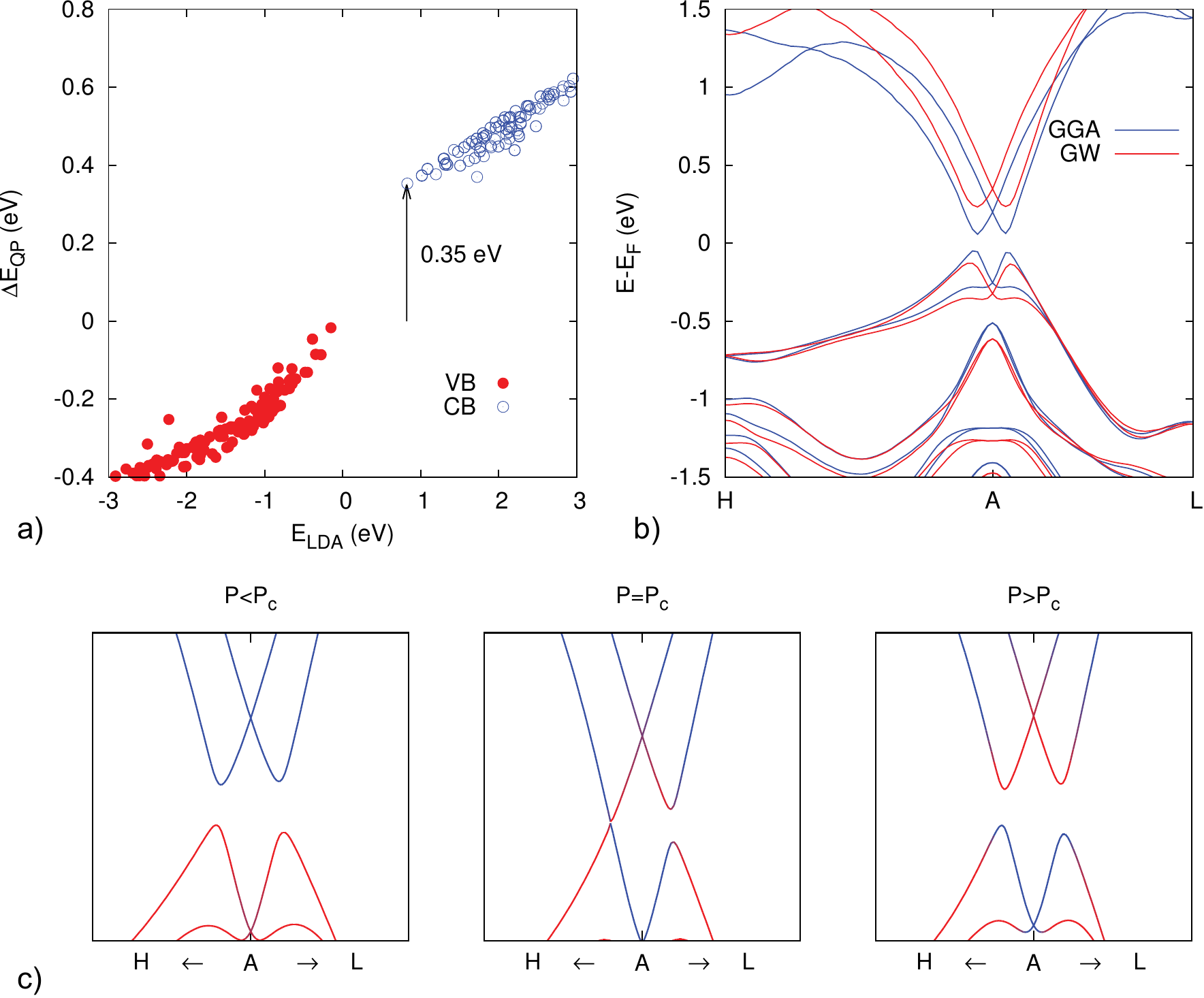}
\caption{(a) Quasiparticle self-energy corrections as
a function of LDA energies for bulk BiTeI,  (b) GGA and GW band structure of bulk BiTeI at ambient pressure, and (c) Topological band structure transition around the critical pressure $P_C$. The color correspond to the weight of the state on the Bi (blue) and Te (red) atoms. }
\label{fig:figs1}
\end{figure}

The atomic positions and cell shape were relaxed at constant volume for a range of volume between $0.85V_0$ and $1.15V_0$, where $V_0$ is the experimental unit cell volume ($V_0=111.76$\AA$^3$). The pressure was obtained by fitting the Murnaghan equation of state to the  energy vs. volume curve. The gap as a function of the pressure is shown on Fig.~\ref{fig:figs2}. A closing of the gap is expected around  $P_C=2.7$GPa. A sketch of the band structure transition is shown on Fig.~\ref{fig:figs1}c). The closing of the gap takes place near the A point along the A$\rightarrow$H line in the Brillouin zone.

The low value of $P_C$ predicted by GGA is due to the underestimation of the band gap inherent to the DFT methodology. For a better estimation of the band gap, we performed a GW calculation. The quasiparticle energies were evaluated within the G$_0$W$_0$ approximation to the electron self-energy starting from non relativistic LDA results for ambient pressure BiTeI using the approach of Hybertsen and Louie~\cite{hyb86}. This first-principles GW methodology is implemented in the BERKELEYGW code~\cite{des12}. The quasiparticle self-energy correction is shown on Fig.~\ref{fig:figs1}a. The correction leads to an opening of the non-relativistic band gap of $0.35$eV.
The ambient pressure quasiparticle correction was applied to the non-relativistic GGA band structure at different volume using the scissor shift approximation. The spin-orbit effect were included as a last step by calculating the spin-orbit matrix element between the scissor-shifted Kohn-Sham GGA eigenstates following the approach of Ref.~\onlinecite{hyb86-2}. The bands structure at ambient pressure obtained from this methodology is shown in Fig.~\ref{fig:figs1}b.
The band gap as a function of the pressure from GW is shown on Fig.~\ref{fig:figs2}. The critical pressure within this approximation is $\sim 10$ GPa.

\section{2. Band structure of the high pressure phase }

\begin{figure}[tb]
\centering
\includegraphics[width=0.8\columnwidth]{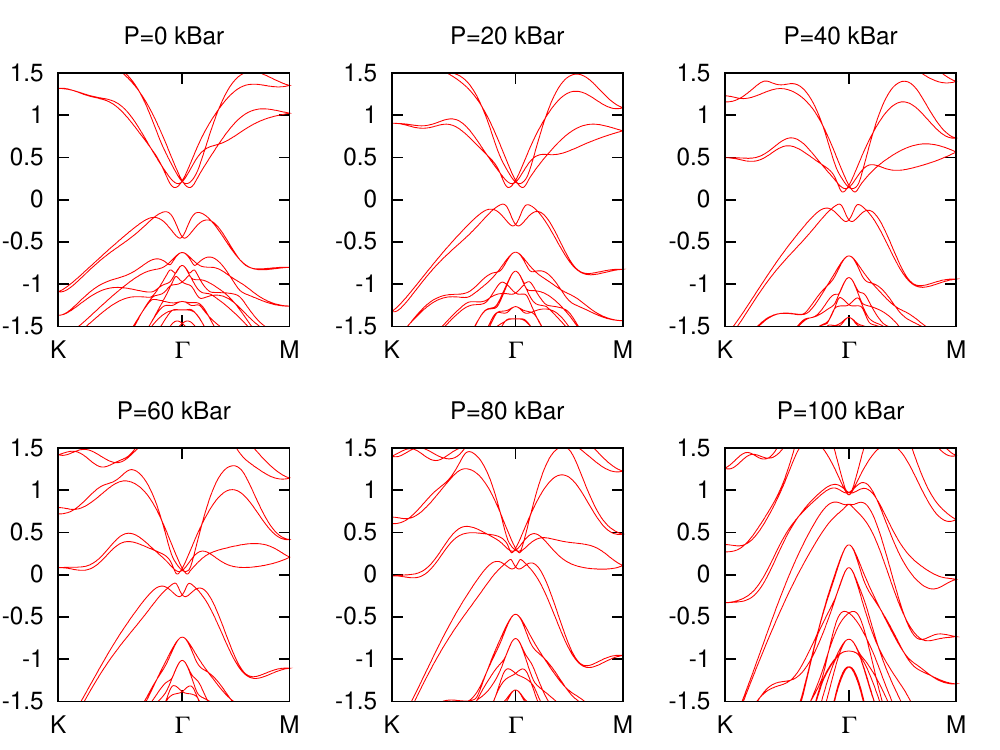}
\caption{Band structure of the P6$_3$mc phase under pressure }
\label{fig:figs3}
\end{figure}

Since the optical properties of BiTeI progress gradually through the crystallographic phase transition (see Fig.~2 in main text), our hypothesis is that the high pressure phase is structurally close to the ambient pressure phase.
A possible candidate for the high pressure phase is the structure of the chemically related compound BiTeCl (space group $P6_3mc$ with 6 atoms per unit cell). This structure correspond to a doubling of the cell in the direction perpendicular to the layers and differs from the ambient pressure phase by the stacking of the two BiTeI layers. We calculated the band structure of this phase at different pressure within the LDA approximation (Fig.~\ref{fig:figs3}). The results show that in this phase the gap also decreases gradually as a function of pressure. However, in contrast with the ambient pressure phase, the $P6_3mc$ phase terminates in a topologically trivial metallic state. Thus no gap reopening or topological phase transition is expected. This is consistent with the gradual closing of the gap observed in experiment.

\section{3. Transmission}

\begin{figure}[htb]
\centering
\includegraphics[width=0.8\columnwidth]{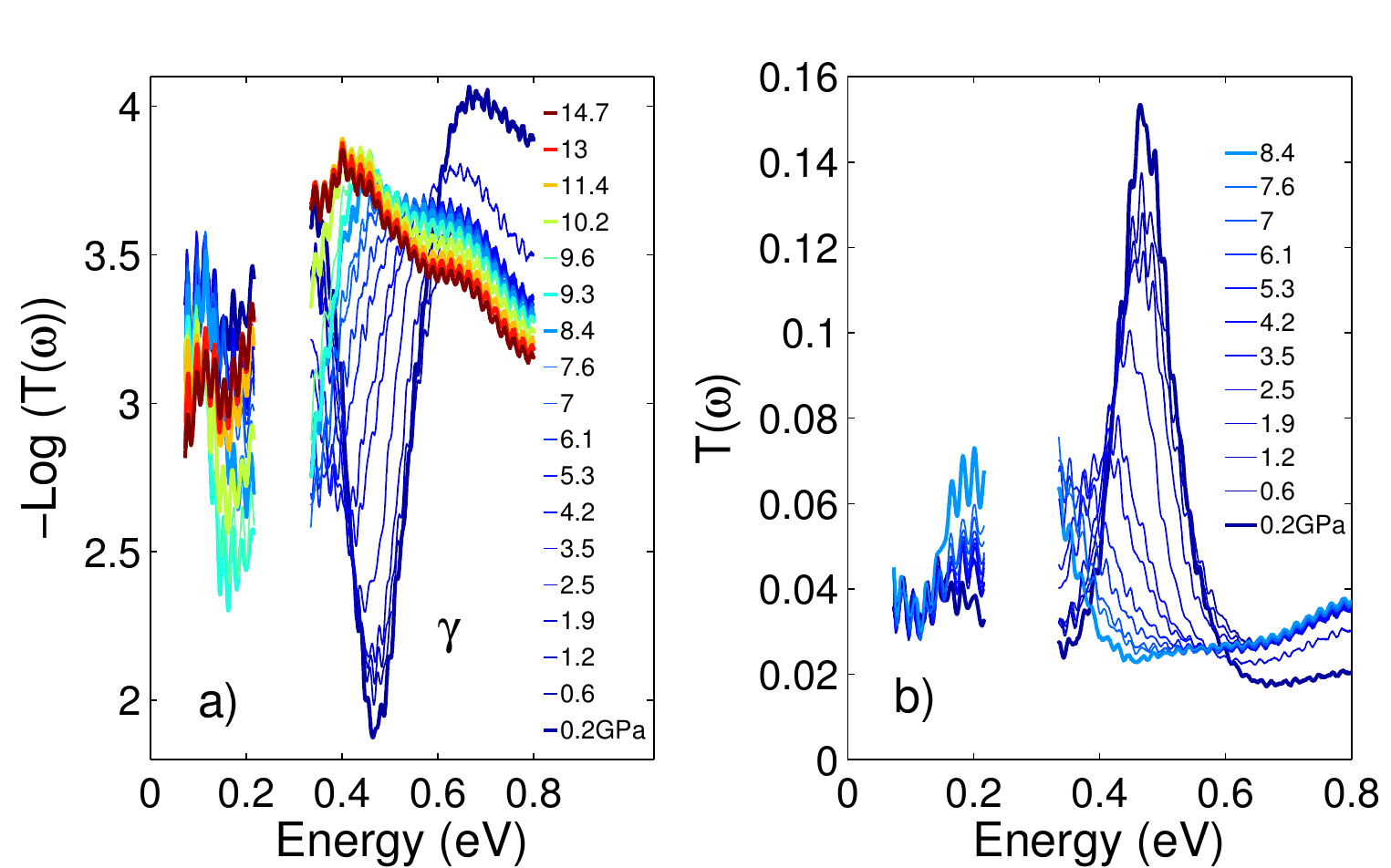}
\caption{Transmission data. a) Conductivity for all pressures and b) Transmission for $p<9$~GPa. Color code matches the one used in Fig.~2b of the manuscript.}
\label{fig:suplT}
\end{figure}

Fig.~\ref{fig:suplT} presents transmission data down to $0.07$~eV. The optical conductivity is given to a good approximation by  $-\log(T(\omega))$, which is shown in Fig.~\ref{fig:suplT}a. Clearly the $\gamma$ feature remains close to its ambient value of $\sim 0.4$~eV even up to $9$~GPa (blue curves), indicating a gradual evolution of the gap as a function of pressure in this pressure range. The lower pressure $p<9$~GPa data displayed in Fig.~\ref{fig:suplT}b shows no increase of transmission at low energy up to $9$~GPa where all curves are superposed.

